\documentclass{llncs}
\pdfoutput=1

\usepackage[T1]{fontenc}
\usepackage[british]{babel}
\usepackage[paperwidth=155mm,paperheight=235mm,left=16.5mm,right=16.5mm,top=16.8mm,bottom=25.2mm]{geometry} 

\makeatletter
\RequirePackage[bookmarks,unicode,colorlinks=true,breaklinks=true]{hyperref}%
   \def\@citecolor{blue}%
   \def\@urlcolor{blue}%
   \def\@linkcolor{blue}%

\def\orcidID#1{\smash{\href{http://orcid.org/#1}{\protect\raisebox{-1.25pt}{\protect\includegraphics{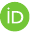}}}}}
\makeatother

\usepackage{cite}
\usepackage[dvipsnames]{xcolor}
\usepackage{graphicx}
\usepackage{xspace}
\usepackage{bbding}

\usepackage{enumitem}
\usepackage{amssymb}
\usepackage{amsmath}
\usepackage{cleveref}

\setitemize{noitemsep,topsep=0pt,parsep=0pt,partopsep=0pt,leftmargin=12.0pt}
\setenumerate{noitemsep,topsep=0pt,parsep=0pt,partopsep=0pt,leftmargin=12.5pt}
\setdescription{noitemsep,topsep=0pt,parsep=0pt,partopsep=0pt,leftmargin=12.5pt}

\newcommand{\eg}{e.g.\ }
\newcommand{\ie}{i.e.\ }

\newcommand{\sunit}[1]{\text{\,#1}}

\begin{document}

\title{%
Reproduction Report for SV-COMP 2023%
\thanks{%
This work was supported by
NWO VENI grant no.\ 639.021.754
%and has received funding from the European Union's Horizon 2020 research and innovation programme under the Marie Sk{\l}odowska-Curie grant agreement No 101008233.
and the EU's Horizon 2020 research and innovation programme under MSCA grant agreement 101008233.
}%
}
\author{
Marcus Gerhold\,\orcidID{0000-0002-2655-9617} \and
Arnd Hartmanns\,\orcidID{0000-0003-3268-8674}
}
\authorrunning{M. Gerhold, A.\ Hartmanns}

\institute{
University of Twente, Enschede, The Netherlands\\
\email{\{\,m.gerhold,\,a.hartmanns\,\}@utwente.nl}}
%\date{\today}
\maketitle

\begin{abstract}
The Competition on Software Verification (SV-COMP) is a large computational experiment benchmarking many different software verification tools on a vast collection of C and Java benchmarks.
Such experimental research should be reproducible by researchers independent from the team that performed the original experiments.
In this reproduction report, we present our recent attempt at reproducing SV-COMP 2023:
We chose a meaningful subset of the competition and re-ran it on the competition organiser's infrastructure, using the scripts and tools provided in the competition's archived artifacts.
We see minor differences in tool scores that appear explainable by the interaction of small runtime fluctuations with the competition's scoring rules, and successfully reproduce the overall ranking within our chosen subset.
Overall, we consider SV-COMP 2023 to be reproducible.
\end{abstract}

\section{Introduction}
\label{sec:Introduction}

The International Competition on Software Verification (SV-COMP) compares software verification tools on a very large amount of benchmark verification tasks.
Associated to the TACAS conference, its first edition took place in 2012~\cite{Bey12}.
This report is about SV-COMP 2023, the competition's 12th edition.
SV-COMP 2023 is described in a competition report~\cite{Bey23} and on its website~\cite{SvComp23Website}.
The competition report provides a summary of the competition setup and presents the overall tool results and winners in different categories.
The website provides the full details about the competition, including its benchmark set and scoring scheme, and detailed plots and tables of the tools' results.
52 different verification tools participated in SV-COMP 2023, which consists of 24,391 different benchmark problems (of which 23,805 are in C and 586 are in Java) in nine categories.

SV-COMP is an example of a large computer science experiment that evaluates many tools on many benchmark instances.
The outcomes of this experiment are used to rank tools in terms of their ability to solve problems correctly, and in terms of their performance concerning runtime and energy usage.
For the outcomes of experimental research to be trustworthy, the experiment needs to be repeatable and reproducible.
Here, following the ACM terminology~\cite{AcmTerminology}, \emph{repeatability} means that the same researchers that performed the original experiment can repeat it, \ie run the same benchmarks again on the same system obtaining the same results (up to stated precision limits, which is particularly relevant for randomised/statistical experiments that \emph{will} show different results on each repetition).
This is a very basic requirement; arguably, non-repeatable experiments should not be published in the first place.
\emph{Reproducibility}, on the other hand, means that
\begin{quote}
  \it The measurement can be obtained with stated precision by a different team using the same measurement procedure, the same measuring system, under the same operating conditions, in the same or a different location on multiple trials. For computational experiments, this means that an independent group can obtain the same result using the author's own artifacts.
  \hfill\cite{AcmTerminology}
\end{quote}
This requires a separate set of researchers, but not necessarily a separate system to run the experiments on.
The EAPLS similarly defines results to have been reproduced so that a ``Results Reproduced'' artifact badge can be awarded if
\begin{quote}
  \it The main results reported in the paper have been obtained in a subsequent study by a person or team other than the author(s), using (in part) artifacts provided by the author(s).
  \hfill\cite{EaplsArtifacts}
\end{quote}
Of note here is that this definition only requires ``the main results'' reported in a paper describing an experimental study to be obtained once more.

Reproducibility is a sign of quality for experimental research; in areas of computer science related to SV-COMP in particular, artifact evaluations associated to conferences such as TACAS (since 2018) encourage and reward reproducible results.
In this report, we summarise our recent attempt at reproducing SV-COMP 2023 and its outcomes.

\section{Reproducing SV-COMP}

To the best of our knowledge, our work is the first documented attempt to reproduce a tool competition of the scale of SV-COMP.
We thus took a practical approach to find out if we can, with limited effort, reproduce enough of SV-COMP 2023 in a sufficiently independent manner to consider the competition as a whole \emph{likely reproducible}.

SV-COMP provides artifacts~\cite[Table~3]{Bey23} that include the participating tools (in binary or source code form), the benchmark instances that they are executed on, and the scripts that were used to run the competition.
Owing to SV-COMP's large size---running SV-COMP 2023 in its entirety once required 1,114 days of CPU time to execute its 490,858 verification runs---the benchmarking is performed on a large cluster of 168 machines administrated by the competition organiser's group at LMU Munich.
Inspecting these artifacts, we find that SV-COMP should clearly be repeatable.

Without comparable resources, however, any attempt at reproducing anything but a very small fraction of SV-COMP within a reasonable amount of time is infeasible; and even if one had enough time to spare, the competition's scripts are closely geared towards its specific execution environment, requiring additional work to adapt them to different settings.
Yet, following the ACM definition, reproducibility does not require the experiments to be in a different setting using a different setup: as long as the team performing the measurements is different, they may use ``the same measurement procedure [and] the same measuring system.''
This is the first point where we apply our practical-with-limited-effort methodology:
The organiser of SV-COMP 2023 granted us access to their cluster, so that we could re-use the same ``measurement procedure'' (scripts and setup) on the same ``measuring system'' (the cluster with the same machines).

Still, running SV-COMP on this cluster originally took the organiser about 8 days (of wall-clock time).
Given that our intention was for this reproduction report to be available together with the competition report, and our unfamiliarity with the infrastructure surely resulting in additional delays, a full reproduction remained infeasible for us despite access to the cluster.
Instead, as the second practical-with-limited-effort point, we selected a subset of the competition to reproduce, as a ``spot check''.
If our reproduction results are close to the original results, and if we made a representative selection, we should be allowed to generalise our reproduction study's outcome to the whole competition.

\subsection{Results to Reproduce}

We consider the main result presented for SV-COMP 2023 in its competition report and on its website to be the scoring and ranking of the regularly participating verification tools~\cite[Table~8]{Bey23}.
For each benchmark instance, SV-COMP uses a soft timeout of 900\sunit{s} after which a tool's result no longer counts for scoring, and a hard timeout of around 960\sunit{s} after which the tool is forcibly terminated.
For every result produced within the soft timeout, the tool receives between $-32$ (for incorrectly reporting ``true'') and $+2$ (for correctly reporting ``true'') points~\cite[Table~1]{Bey23}.
In this way, the actual tool runtime---although collected and presented in the detailed results tables---does not matter as long as it is within the timeout and except in case of a tie in score.
Similar to the timeout, the memory available to tools was limited to 15\sunit{GB}.
Our reproduction attempt thus seeks to check whether we obtain the same \emph{ranking}, checking the scores to see how much (if any) deviation we observe.

In particular, given its evaluation scheme, the main results of SV-COMP should be resilient to small changes in runtime, unless the time a tool needs for a certain benchmark instance is right on the timeout boundary.
However, SV-COMP would not be resilient to highly nondeterministic tools that have vast runtime or memory usage differences between executions with the same inputs, and to tools that are significantly nondeterministic in their results (which would arguably be tools of limited usefulness, thus tool authors are expected to avoid such behaviour---so that it would indicate tool bugs).

\subsection{Selected Subset of Experiments}
\label{sec:Subset}

The subset we selected to reproduce SV-COMP's main results has two parts.

\subsubsection{Ranking check.}

First, we spot-check the regular verification tools ranking presented in the competition report (explicitly listed for the top three tools~\cite[Table~10]{Bey23} and implicitly given via the total scores otherwise~\cite[Table~8]{Bey23}):

\begin{itemize}
\item
Re-run category \textit{\bfseries ConcurrencySafety} for ranking places 2-4 (tools \textit{UAutomizer}, \textit{UGemCutter}, and \textit{UTaipan}, respectively) because the scores are close (being $2717$, $2710$, and $2612$, respectively).
\item
Re-run category \textit{\bfseries SoftwareSystems} for ranking places 1-2 (tools \textit{Symbiotic} and \textit{Bubaak}, respectively), again since they are close in scores (of $1604$ and $1589$, respectively), and because they produce one incorrect result each.
\item
Re-run category \textit{\bfseries JavaOverall} for ranking places 1-3 (tools \textit{JBMC}, \textit{GDart}, and \textit{MLB}, respectively) to check that the Java-based part of the competition (which is much smaller---just one category---than its C-based part) is in order.
\end{itemize}

\subsubsection{Tools check.}

Second, we spot-check specific tools in specific categories that showed different interesting behaviour or characteristics:

\begin{itemize}
\item
Re-run tool \textit{\bfseries VeriFuzz} in categories \textit{NoOverflows} and \textit{Termination} because it has one quite negative score in one category ($-500$ in \textit{NoOverflows}) while winning a gold medal (first place) in the other (with score $2305$).
\item
Re-run tool \textit{\bfseries Symbiotic} in categories \textit{MemSafety} and \textit{Termination} because it is a ``portfolio'' tool, where the use, selection, or ordering of the multiple algorithms could lead to nondeterministic behaviour.
\item
Re-run tools \textit{\bfseries LF-checker} and \textit{\bfseries Deagle} in category \textit{ConcurrencySafety} because they participate in only this category, and one of them (\textit{Deagle}) wins the gold medal there.
\end{itemize}

\bigskip\noindent
All scores we mention above are normalised category scores as described at \href{https://sv-comp.sosy-lab.org/2023/rules.php#scores}{sv-comp.sosy-lab.org/2023/rules.php}.
They are computed from the raw scores of several sub-categories.
In \Cref{sec:ResultsReproduced}, we report raw scores instead, as they are the ones shown on the detailed per-tool tables; as long as the relationship between raw scores remains the same, the overall ranking will not change.
However, smaller differences in raw scores will have a larger impact on the normalised category scores in smaller categories.

\section{Reproduction Results}
\label{sec:Results}

After having identified the main results we want to check, and the subset of experiments to use for this purpose, we started our reproduction attempt.
The artifacts for SV-COMP contain a readme file describing the organisation of the included data (such as per-tool benchmark results), which covers all the data generated in SV-COMP 2023 and processed to produce the main results.
However, they do not contain detailed instructions for reproducing the competition.
In addition to granting us access to their cluster, the competition's organiser thus also provided such instructions to us.
In the subsequent reproduction process, we made observations about the reproduction process, and about the reproduction of the competition's main results.

\subsection{Reproduction Process}

In the process of re-executing our selected subset, we encountered several small problems in following the instructions.
For example,

\begin{itemize}
\item
a few required Python dependencies were not installed, and installing them was not part of the first version of our instructions;
\item 
our instructions were intended for reproducing one sub-category at a time, but not an entire category in one go---we then received expanded instructions for how to assemble the necessary parameters for entire categories; and
\item
creating tables according to our instructions at first listed results only as ``correct'' or ``incorrect'', but not also as ``correct-unconfirmed'' as in the official result tables, leading to significantly different scores---this was because the organiser had assumed we would not want to run the results validation procedure, for which we then received expanded instructions.
\end{itemize}
Overall, this was a learning process both for us as well as for the competition organiser:
We increasingly understood the competition's setup, and the organiser gained an understanding of what level of documentation and tooling is necessary to support a smooth reproduction.
In particular, all information and data was in principle available from the start: We could have studied the shell scripts, tools, folder structure, etc.\ that were part of the artifacts in detail and thereby reverse-engineered the entire process.
This however would not have been feasible given the limited time we had, and in general would make independent reproduction hard and unlikely to happen.

Overall, though, the problems we encountered during the reproduction process actually increased our confidence in the soundness of the competition setup and its artifacts:
By running scripts in unintended ways, deviating from the original instructions, and creating our own variants of \eg the table templates, we tested the flexibility of the artifacts and ensured that they are reusable.
In particular, it would be very unlikely for artifacts exercised in this way to merely ``simulate'' running the competition and delivering the desired results---it rather looks like we indeed performed the experiments that SV-COMP claims to have performed once again!

\subsection{Reproduction of the Main Results}
\label{sec:ResultsReproduced}

The result tables for our selected subset of (as described in \Cref{sec:Subset}) are available at \href{https://arnd.hartmanns.name/sv-comp-2023-repro/}{arnd.hartmanns.name/sv-comp-2023-repro}.
Overall, the results we obtained are in line with those of SV-COMP 2023, with small deviations in scores throughout but no change in ranking.

\subsubsection{Ranking check.}
We first look at our spot-check of the ranking in 3 categories.
In category \textit{\bfseries ConcurrencySafety}, we find a small increase in scores compared to the original SV-COMP results for \textit{UAutomizer} (from 2725 to 2733) and \textit{UTaipan} (from 2607 to 2613).
With \textit{UGemCutter} reproducibly at 2714 and the gold-medal winner \textit{Deagle} at 4754 in SV-COMP, the ranking remains unchanged.
However, especially \textit{UAutomizer} and \textit{UGemCutter} are very close (score difference of $11$), and the changes in scores (of $+8$ and $+6$) are about on the same order of magnitude as the differences between the tools' scores here.
The same happens in category \textit{\bfseries JavaOverall}, though with smaller absolute differences given the smaller category (with \textit{JBMC} going from 669 to 667 and \textit{MLB} from 495 to 496).
In \textit{\bfseries SoftwareSystems}, scores and correct/incorrect result numbers match exactly.

\subsubsection{Tools check.}
For the individual tools, we confirmed the negative result of \textit{\bfseries VeriFuzz} in \textit{NoOverflows}, albeit with a small improvement (from $-87$ to $-80$).
In the \textit{Termination} category, something went wrong in our reproduction:
We obtained the same number of ``correct true'' results, but not a single ``correct false'' result; and also the distinction between ``correct'' and ``correct-unconfirmed'' is missing in our results table, despite the validation clearly having worked for \textit{VeriFuzz} in the \textit{NoOverflows} category.
These differences look more like a bug in the scripts or an error on our side than a failure in reproducibility related to the tool or its execution; we are currently investigating what the root problem is.
For \textit{\bfseries Symbiotic} and \textit{\bfseries Deagle}, we obtained exactly matching scores, and small differences only for \textit{\bfseries LF-checker}.

\bigskip\noindent
In terms of the secondary characteristics like runtime, we only saw small changes throughout all our experiments.
We looked into the raw results data (in the corresponding \texttt{.csv} files) for some of the cases of slightly different scores.
We found various types of differences that appear reasonable overall.
For example, in the case of \textit{UAutomizer},
\begin{itemize}
\item
several benchmark instances changed from timeout to out-of-memory and vice-versa---which is reasonable for difficult instances where the tool needs or tries to use all available resources; and
\item
some changed between a timeout with a result and a pure timeout---which means that the tool once ran into the soft and once into the hard timeout, showing executions at the boundary of the runtime budget that make it or not due to small fluctuations.
\end{itemize}

\section{Conclusion}

We, as researchers independent of the organiser of SV-COMP 2023, were able to re-run a manually but carefully selected subset of SV-COMP 2023 using the organiser's setup and infrastructure.
Our reproduction results show small differences in scores, which are mostly well-explained due to occurring for benchmark instances that are barely (not) feasible.
These differences do not change the ranking of tools, which we consider the main result of SV-COMP 2023.
Thus:
\begin{quote}
  Based on a spot-check of a subset of its experiments, using the same experimental setup and environment, we consider \textbf{the main results of SV-COMP 2023 to be reproducible}.
\end{quote}
However, the fluctuations we see combined with the closeness of some of the tools' scores in some categories should act as a warning to the SV-COMP organiser to consider the fairness of the competition's ranking in such close calls.

We also found that SV-COMP is currently not set up for ``easy'' reproduction: 
While all the material is available, we needed to obtain instructions from the competition organiser, which we had to get updates for in an iterative process whenever we found a bug in the instructions or encountered an unforeseen situation.
This however increased our confidence in the ``honesty'' of the SV-COMP artifacts, and provided valuable insights to the organiser for easing the reproducibility of future editions of SV-COMP.

Finally, an important consideration is whether a spot-check-based approach like ours is useful or sufficient to establish whether an extensive experiment like SV-COMP is reproducible, or has successfully been reproduced.
Given the extent of SV-COMP, a full reproduction is a significant time investment with access to the competition's cluster infrastructure, and likely not feasible without.
Although we put thought into our selection of the subset, we could naturally have missed a highly nondeterministic tool that just happened to win a medal by chance.
We stipulate that an approach using a \emph{randomly sampled} subset of experiments could result in a more formal, albeit statistical, guarantee.

\paragraph{Data availability.}
%A dataset to replicate the experimental evaluation, including the exact versions of the tools and models used, is archived and available at DOI \href{https://doi.org/TODO}{TODO}~\cite{PaperArtifact}.
The tables of results that we reproduced as described in this report are available at \href{https://arnd.hartmanns.name/sv-comp-2023-repro/}{arnd.hartmanns.name/sv-comp-2023-repro}~\cite{PaperArtifact}.

\bibliography{paper}
\bibliographystyle{splncs04}

\end{document}